\documentclass[sn-nature]{sn-jnl}

\usepackage{graphicx}%
\usepackage{multirow}%
\usepackage{amsmath,amssymb,amsfonts}%
\usepackage{amsthm}%
\usepackage{mathrsfs}%
\usepackage[title]{appendix}%
\usepackage[table]{xcolor}
\usepackage{textcomp}%
\usepackage{manyfoot}%
\usepackage{booktabs}%
\usepackage{algorithm}%
\usepackage{algorithmicx}%
\usepackage{algpseudocode}%
\usepackage{listings}%
\usepackage{float}                
\usepackage{subfig} 
\usepackage{enumitem}
\usepackage{bm}
\usepackage{array}
\usepackage{rotating}
\usepackage{tabularx}
\usepackage{hyperref}
\usepackage{setspace}
\usepackage{lineno} 

\begin{document}

\setstretch{1.5}

\title[]{Spatio-temporal Patterns between ENSO and Weather-related Power Outages in the Continental United States}

\author[1]{\fnm{Long} \sur{Huo}}

\author*[1]{\fnm{Xin} \sur{Chen}}\email{xin.chen.nj@xjtu.edu.cn}

\author[2, 3]{\fnm{Kaiwen} \sur{Li}}

\author*[4]{\fnm{Fengying} \sur{Cai}}\email{fenyingc@pik-potsdam.de}

\author[4, 5]{\fnm{J\"urgen} \sur{Kurths}}

\affil[1]{\orgdiv{Center of Nanomaterials for Renewable Energy, State Key Laboratory of Electrical Insulation and Power Equipment, School of Electrical Engineering}, \orgname{Xi'an Jiaotong University}, \city{Xi'an}, \postcode{710054}, \country{China}}

\affil[2]{\orgname{School of National Safety and Emergency Management, Beijing Normal University}, \postcode{100875} \city{Beijing}, \country{China}}

\affil[3]{\orgname{Faculty of Geographical Science, Beijing Normal University}, \postcode{100875} \city{Beijing}, \country{China}}

\affil[4]{\orgname{Potsdam Institute for Climate Impact Research (PIK), Member of the Leibniz Association}, \postcode{14473} \city{Potsdam}, \country{Germany}}

\affil[5]{\orgname{Department of Physics, Humboldt University at Berlin}, \postcode{12489} \city{Berlin}, \country{Germany}}


\abstract{El Niño-Southern Oscillation (ENSO) exhibits significant impacts on the frequency of extreme weather events and its socio-economic implications prevail on a global scale. However, a fundamental gap still exists in understanding the relationship between the ENSO and weather-related power outages in the continental United States. Through 24-year (2000-2023) composite and statistical analysis, our study reveals that higher power outage numbers (PONs) are observed from the developing winter to the decaying summer of La Niña phases. In particular, during the decaying spring, high La Niña intensity favors the occurrences of power outage over the west coast and east of the United States, by modulating the frequency of extreme precipitations and heatwaves. Furthermore, projected increasing heatwaves from the Coupled Model Intercomparison Project Phase 6 (CMIP6) indicate that spring-time PONs over the eastern United States occur about 11 times higher for the mid-term future (2041–2060) and almost 26 times higher for the long-term future (2081–2100), compared with 2000–2023. Our study provides a strong recommendation for building a more climate-resilient power system.}

\keywords{Power outage, El Niño-Southern Oscillation (ENSO), Extreme weather}

\maketitle

\section*{Introduction}
Under the background of anthropogenic climate change, increasing occurrences of extreme weather events have posed a profound threat to the resilience of energy infrastructures \cite{bib36, bib37, bib38, bib39, bib40}. The frequency of climate-induced energy disruptions has been growing continuously over the past two decades, particularly evidenced by outages in power systems. Between 2000 and 2021, approximately 83\% of the United States (U.S.) major power outages are attributed to weather-related events, with a surging trend of average annual power outage numbers (PONs) by about 78\% during 2011-2021, compared to 2000-2010 \cite{bib21}. Serious power outages could be triggered by various extreme weather events such as the 2009 Brazil and Paraguay extreme precipitation and strong wind \cite{bib32}, the 2020 California heatwave \cite{bib33}, and the 2021 Texas winter storm \cite{bib34}, causing millions of residents without power supply for hours to days, billions of dollars economic costs, and even hundreds of human deaths. It is therefore imperative to deepen our understanding of the ramifications between climate variability and weather-related power outages to establish a climate-resilient power system.

For weather-related power outages, previous studies primarily concentrated on power outage modeling techniques \cite{bib54, bib55, bib56}, specific weather patterns (or compound hazards) \cite{bib41, bib42, bib43, bib44, bib45}, assessment of socio-economic and health impacts \cite{bib46, bib47, bib48, bib49}, and mitigation strategies and forecasting methods \cite{bib50, bib51, bib52, bib53}. However, little is known about how the power outage is modulated by the well-recognized climate variability, such as El Niño-Southern Oscillation (ENSO). 

It is broadly acknowledged that ENSO is one of the most pronounced climate phenomena, and the ENSO-induced atmospheric teleconnection could modulate the frequency and intensity of extreme weather events including extreme rainfall and heatwaves across the globe \cite{bib57, bib58}. El Niño and La Niña refer to the warm and cold phases of ENSO, respectively, with a recurrence of about two to seven years. The impacts of ENSO on power systems could be summarized via two approaches. On the one hand, gradually raising air and sea surface temperature variability are closely linked to ENSO \cite{bib59, bib60}, which are responsible for the surging cooling demand \cite{bib61} and decreasing thermoelectric generation usable capacity \cite{bib62}. On the other hand, extreme ENSO phases favor the occurrences of extreme weather events, $i.e.$, extreme precipitation \cite{bib64}, regional flooding \cite{bib67}, and severe cold snap \cite{bib66}, possibly leading to electrical infrastructure performance degradation or even damages \cite{bib65, bib32, bib34}. The impacts of ENSO are potential leading origins of power outages. Nevertheless, the nature of the linkage between ENSO and weather-related power outages remains largely unexplored.

Here, via a thorough statistical analysis of the historical U.S. power outage records, ENSO monitoring index, and $in$ $situ$ observation-based weather data (see Methods for details)  spanning from 2000 to 2023, we reveal a nexus among ENSO, local extreme weather events, and power outages. In particular, La Niña is found to markedly encourage the growing PONs, which is most significant during the decaying spring, followed by decaying summer and simultaneous winter, while insignificant during decaying autumn. High La Niña intensity intensifies PONs by modulating U.S. extreme weather events including heatwaves, cold snaps, and extreme precipitation among the western, southern, and mid-eastern U.S. Furthermore, future changes in heatwave-related PONs will be assessed by using heatwave projections from the Coupled Model Intercomparison Project Phase 6 (CMIP6), and the historical quantitative regression of heatwave-related PONs.

\section*{Statistical relationship between the ENSO and power outages}
\begin{figure}[htb]
	\centering
	\includegraphics[width=1.2\textwidth]{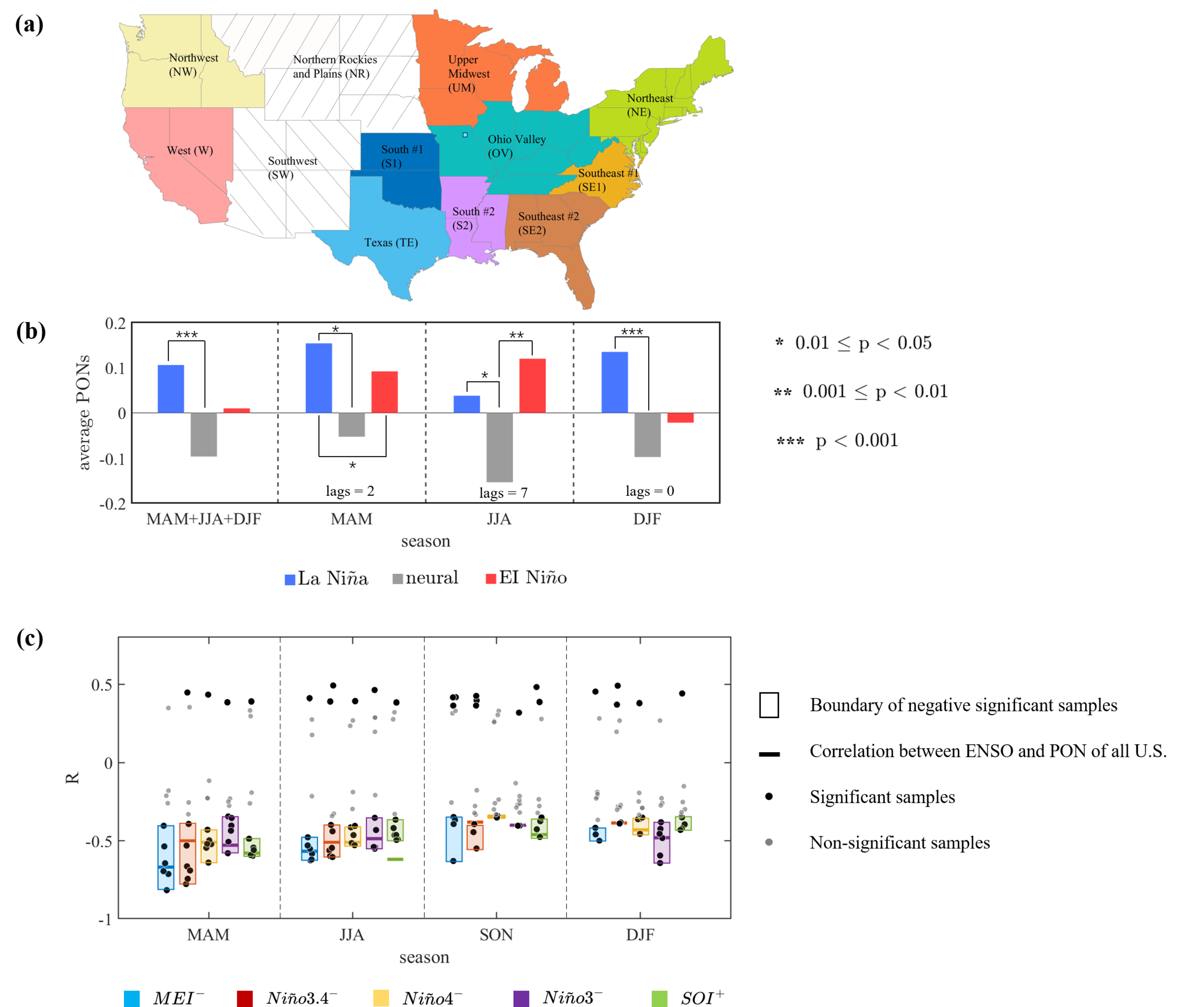}
	\caption{\textbf{Statistical relationship between the EI Ni\~{n}o-Southern Oscillation (ENSO) and power outages}. (a) Climate regions of the U.S. There are ten climate regions. Note that two climate regions (NR and SW) are not analyzed in this study due to limited samples of weather-related power outages ($<$ 20 from 2000 to 2023). (b) Composite seasonal average power outage numbers (PONs) of all U.S. during La Ni\~{n}a, EI Ni\~{n}o, and neural-ENSO phases. (c) Cross-correlations between the La Niña intensities based on different ENSO monitoring indices (MEI, SOI, Ni\~{n}o 3.4, Ni\~{n}o 3, and Ni\~{n}o 4) and the regional PONs for each season. The samples with significantly (non-significantly, $p>0.05$) cross-correlations are indicated by black (gray) dots. In (B) and (C), the acronyms MAM, JJA, SON, and DJF refer to spring (March, April, May), summer (June, July, August), autumn (September, October, November), and winter (December, January, February), respectively.  The ENSO phase associated with a power outage in spring, summer, autumn, or winter is determined by a $\pm$0.5 threshold of the MEI index 3, 6, 9 or 0 months (with ± 1-month tolerance interval) before the power outage occurs, $i.e.$, if the MEI $<$ -0.5, it indicates the La Niña phase; $>$ +0.5, it's the El Niño phase; otherwise, it's the neutral-ENSO phase. }\label{fig1}
\end{figure}

What is the underlying indication between the ENSO phenomenon and power outages?
To address this question, a statistical analysis is conducted between the historical power outage records and climate data within the continental U.S. For a refined analysis granularity, the nine climate regions identified by National Centers for Environmental Information scientists \cite{bib20} are augmented into twelve, including Northwest (NW), West (W), Northern Rockies and Plains (NR), Southwest (SW), South \#1 (S1), South \#2 (S2), Texas (TE), Upper Midwest (UM), Ohio Valley (OV), Southeast \#1 (SE1), Southeast \#2 (SE2), and Northeast (NE) (Fig. \ref{fig1}a). We aggregate the PONs from all states in each climate region to obtain the regional PONs. Note that two regions (NR and SW) are excluded from the analytical scope due to limited samples of weather-related power outages ($<$ 20 from 2000 to 2023). In addition, both the power outage and climate data are processed with data preprocessing (see Methods for details).

Significant above-normal PON anomalies are observed during the La Niña and El Niño phases compared with the neutral-ENSO phase (Fig. \ref{fig1}b, examined by the ANOVA and Tukey's HSD test, see Methods for details). The MEI index classifies ENSO phases by a $\pm$0.5 threshold, $i.e.$, the months with MEI $<$ -0.5 ($>$ +0.5) refer to the La Niña (El Niño) phase, while other months are the neural-ENSO phase \cite{bib63}. Typically, an ENSO event peaks in the winter and remains exerts a significant impact on the global climate in the decaying year. Therefore, we focus on the potential impact of ENSO on PONs during the simultaneous winter, decaying spring, summer, and autumn.  In particular, for spring, summer, autumn, and winter, we consider the corresponding PON anomalies 3, 6, 9, and 0 months (with ± 1-month tolerance interval, Supplementary Fig. 1) after the ENSO events peaks in winter. Fig. \ref{fig1}b shows that in spring, summer, and winter, as well as the three seasons combined, the seasonal average PONs of all U.S. during the La Niña phase are significantly larger (p $<$ 0.05) compared with the neutral-ENSO phase. The result for autumn is not presented due to the insignificant difference in PONs during the non-neutral and neural phases.
Although in summer, the PON during the El Niño is also notably larger, this does not alter the fact that most power outages happen during the La Niña for spring, summer, and winter. Henceforth, the modulations of La Niña on power outage are emphasized in this study.

To identify the intensity of La Niña, five monthly ENSO monitoring indices (MEI, Niño3.4, Niño3, Niño4, and SOI, see Methods for details) are adopted as proxies. Generally, the La Niña intensity can be measured by the negative anomalies of the MEI, Niño3.4, Niño3, and Niño4, or the positive anomalies of the SOI (denoted as $MEI^{-}$, Niño3.4$^{-}$, Niño3$^{-}$, Niño4$^{-}$, and $SOI^{+}$). The $MEI^{-}$ is also denoted as $Negat$-$MEI$ intensity in the remaining text. 
The time-delayed cross-correlations (CCs) are evaluated between $MEI^{-}$, Niño3.4$^{-}$, Niño3$^{-}$, Niño4$^{-}$, $SOI^{+}$, and the corresponding PONs over ten climate regions(see Methods for details). We reveal that except during autumn, at least one of the five indices shows significantly negative CCs with the PONs in more than half of all ten climate regions (Fig. \ref{fig1}c).
In other words, for most climate regions, the stronger the La Niña, the significantly more PONs.The most evident correlations are detected in the spring, $i.e.$, CCs between $Negat$-$MEI$ intensity and PONs proved to be statistically significant (p $<$ 0.05) and uniformly negative in six climate regions, ranging from $R$=-0.4 to -0.82 ($R$=-0.68 for PON over all U.S.). In addition, during summer, the negative CCs between $Negat$-$MEI$ intensity and PONs ranging from $R$=-0.47 to -0.62 are significant for six climate regions (R=-0.59 for the PON over all U.S.). Winter is featured by consistently negative CCs between Ni\~{n}o3$^{-}$ and PONs in six climate regions ($R$=-0.4 to -0.7 for PONs of climate regions, and $R$=-0.5 for the PON of all U.S.). During autumn, neither significantly negative nor positive CCs prevail in more than half of all climate regions, implying that La Niña during autumn has no consistent spatial trend of increasing PONs.

\section*{Spatio-temporal patterns of weather-related power outages}
Although significant correlations are observed between the PONs during La Niña phases and ENSO monitoring indices, the corresponding underlying mechanism remains unclear. In this regard, we quantify the potential occurrences of extreme weather events as the mediators between La Niña and weather-related power outages. The $in$ $situ$ observation-based weather data (see Methods for details) is used to investigate whether plausible spatio-temporal correlation patterns exist among La Niña, local extreme weather, and the PONs across different seasons and regions. 

\begin{figure}[htb]
	\centering
    \includegraphics[width=0.95\textwidth]{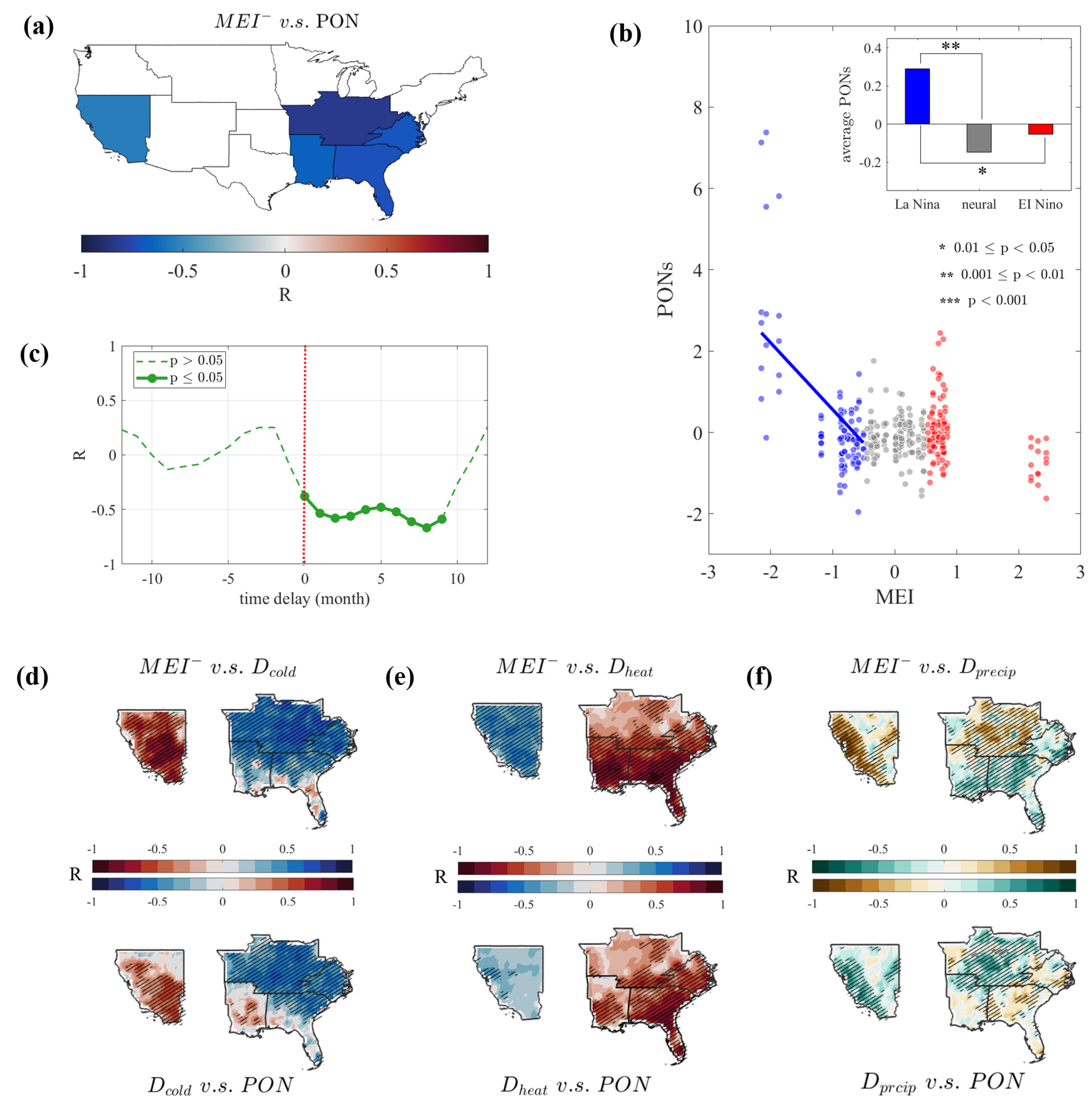}
    \caption{\textbf{Spatio-temporal patterns of weather-related power outages in spring}. (a) Spatial distribution of maximum cross-correlations (CCs) between MEI-based La Ni\~{n}a ($MEI^{-}$) intensity and regional PONs. (b) Scatter plot of $MEI$ and regional PONs. The dots in blue, red, and gray represent samples during La Ni\~{n}a, EI Ni\~{n}o, and neural-ENSO phases. As shown by the blue solid line, the linear regression is significantly (p $\le$ 0.05) negative for $MEI$ $<$ -0.5. The subplot in (b) shows the average PONs during La Ni\~{n}a, EI Ni\~{n}o, and neural-ENSO phases. (c) Time delay CCs between $MEI^{-}$ and PONs of all U.S. The positive (negative) time delays stand for La Ni\~{n}a (PON) predates PON (La Ni\~{n}a). In (d-f), the first row represents maximum CCs between $MEI^{-}$ and frequencies of cold snap $D_{cold}$, heatwave $D_{heat}$, and extreme precipitation $D_{precip}$. The second row is maximum CCs between $D_{cold}$, $D_{heat}$, or $D_{precip}$ and PONs. The shaded areas indicate areas with significant CCs (p $<$ 0.05).}\label{fig2} 
\end{figure}

It is found that during spring, CCs between the $Negat$-$MEI$ intensity and PONs are significantly negative across five climate regions (OV, SE1, SE2, S2, and W) located at the west, south, and mid-east (Fig. \ref{fig2}a). According to the MEI-based ENSO phase classification criterion \cite{bib63}, power outage events are labeled with their corresponding ENSO phase. A significantly (p $<$ 0.05) negative anomaly in PONs is observed during the La Niña phase, while insignificant trends during EI Niño and neural-ENSO phases (Fig. \ref{fig2}b). Fig. \ref{fig2}a together with Fig. \ref{fig2}b imply that more power outages are prone to occur during the stronger La Niña phase. Indeed, we find that the average PON is significantly larger during the La Niña phase compared with the El Niño and neutral phases (subplot of Fig. \ref{fig2}b). 
Moreover, as shown in Fig. \ref{fig2}c, the time-delayed CCs between the $Negat$-$MEI$ intensity and PONs of all U.S. are significant and larger with only non-negative time delays (0 to 9 months). Positive (negative) delays denote that La Niña precedes (lags) power outages. Hence, La Niña always occurs before (or concurrent) with power outages, potentially suggesting that La Niña serves as a dominant factor leading to power outages.

We further discover that La Niña affects PONs by modulating local extreme weathers across regions, including cold snaps, heatwaves, and extreme precipitations (Figs. \ref{fig2}d-f). This could be further explained by the consistency of a pair of CCs, where one is the CC between $Negat$-$MEI$ intensity and frequencies of extreme weathers (top row of Figs. \ref{fig2}d-f), the other is the CC between frequencies of extreme weathers and PONs (bottom row of Figs. \ref{fig2}d-f). The regions featured by the significantly negative CC between $Negat$-$MEI$ intensity and frequencies of extreme weather, and meanwhile, the significantly positive CC between frequencies of extreme weather events and PONs are important. La Nina increases the frequency of extreme weather events in these regions, subsequently causing more power outages. Three regions with corresponding extreme weather events can be found: 1) the western U.S. with cold snap and extreme precipitation, 2) the southeastern U.S. with heatwave, and 3) the mid-eastern U.S. with extreme precipitation. Heatwaves may lead to higher river water temperatures, raising the vulnerability of thermoelectric power supply in the southeastern U.S. \cite{bib72}. The cold snap, heatwave, and extreme precipitations may coincide to become compound events \cite{bib73, bib74}. For the western and southeastern U.S., the storm is a common natural disaster wreaking havoc in coastal areas, usually accompanied by compound weather events like cold snaps and extreme precipitations. For example, in the early spring of 2018, a severe storm swept through California, bringing heavy snow, cold snap, and strong winds, culminating in power outages affecting 10,898 customers and 38MW demand loss \cite{bib1}. In March of 2023, hundreds of tornado outbreaks accompanied by extreme rainfall and strong winds hit more than 10 states across the southeastern and mid-eastern U.S. \cite{bib75}, resulting in about 1 million customers without power supplies \cite{bib1}.

\begin{figure}[htb]
	\centering
    \includegraphics[width=1.1\textwidth]{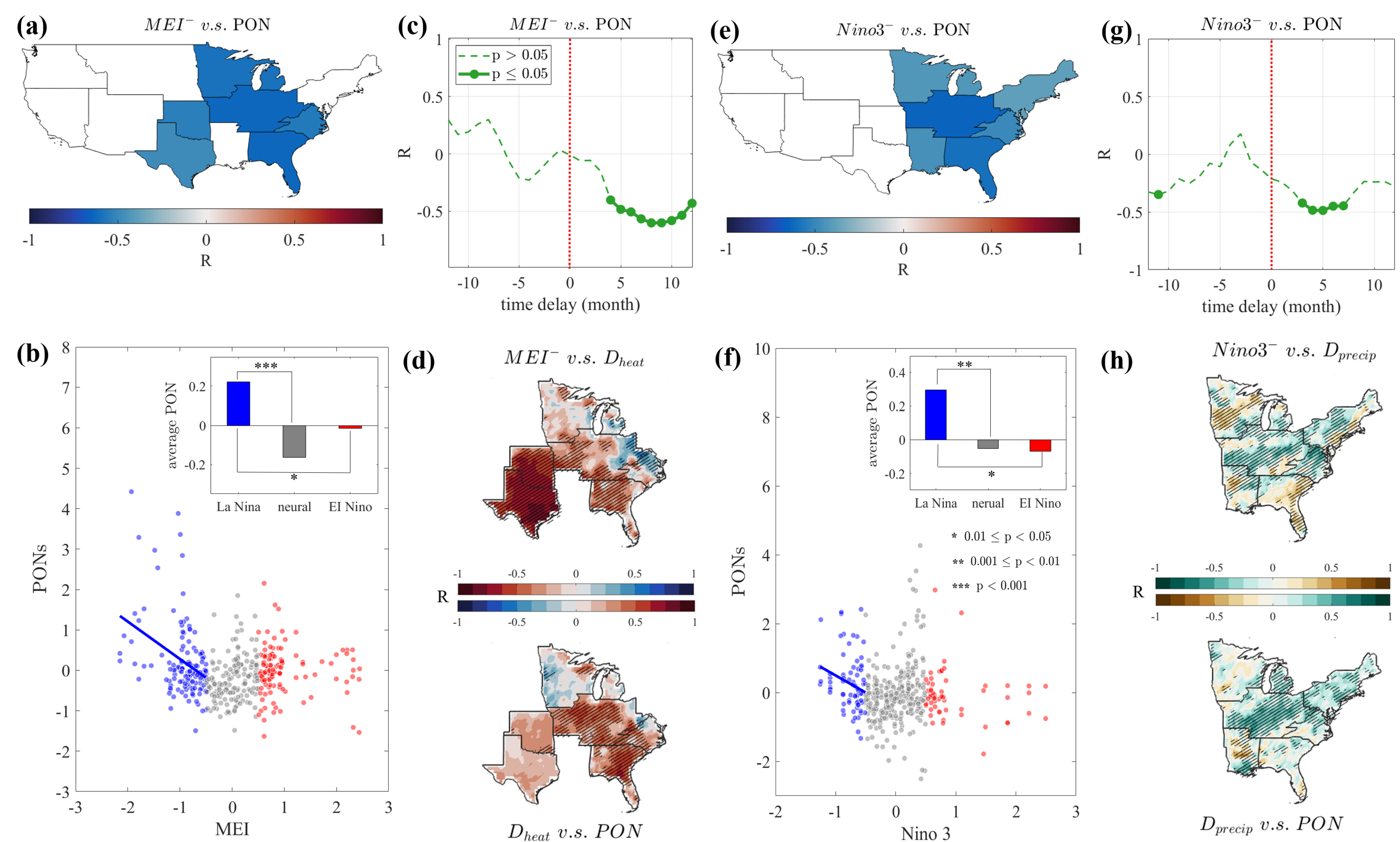}
   \caption{\textbf{Spatio-temporal patterns of weather-related power outages in summer (a-d) and winter (e-h)}. Spatial distribution of maximum cross-correlations (CCs) between MEI-based La Ni\~{n}a ($MEI^{-}$) intensity and regional PONs in summer (a) and maximum CCs between Ni\~{n}o 3-based La Ni\~{n}a (Ni\~{n}o 3$^{-}$) intensity and regional PONs in winter (e). (b) Scatter plot of $MEI$ and regional PONs in summer. (f) Scatter plot of Ni\~{n}o 3 and regional PONs in winter. The dots in blue, red, and gray represent grouped samples in La Ni\~{n}a, EI Ni\~{n}o, and neural-ENSO phases, respectively. Subplots in (b) and (f) show the average PONs during La Ni\~{n}a, EI Ni\~{n}o, and neural-ENSO phases. (c) Time delay CCs between PONs of all U.S. and $MEI^{-}$ during summer. (g) Time delay CCs between PONs of all U.S. and Ni\~{n}o 3$^{-}$ during winter. The positive (negative) time delays stand for La Ni\~{n}a (PON) occurs before PON (La Ni\~{n}a). (d) Maximum CCs between $MEI^{-}$ and heatwave frequencies $D_{heat}$ (top) and maximum CCs between heatwave frequencies and PONs in summer (down). (h) Maximum CCs between Ni\~{n}o 3$^{-}$ and extreme precipitation frequencies $D_{precip}$ (top) and maximum CCs between extreme precipitation frequencies and PONs in winter (down). The shaded areas denote areas with significant CCs (p $<$ 0.05).}\label{fig3}
\end{figure}

In addition, PONs during summer and winter also exhibit conspicuous spatio-temporal patterns with the La Niña and extreme weather. During summer, negative CCs between PONs and $Negat$-$MEI$ intensity are significant in six climate regions (UM, OV, S1, TE, SE1, and SE2) in the south and eastern U.S. (Fig. \ref{fig3}a).  Similarly, negative CCs are observed during winter between the Ni\~{n}o3$^{-}$ and PONs across six climate regions (UM, OV, S2, NE, SE1, and SE2) in the south and eastern U.S. (Fig.\ref{fig3}e). Scatter plots further demonstrate that, for both summer and winter, PONs exhibit negative correlations with corresponding ENSO monitoring indices during the La Niña phase (Figs. \ref{fig3}b and f). Moreover, the PONs during the La Niña phase significantly surpass those during the El Niño and neutral phase (subplots in Figs. \ref{fig3}b and f). These in all suggest that the stronger La Niña probably leads to more PONs during the simultaneous winter and subsequent summer. It is also emphasized that the correlation between La Niña and PONs is unidirectional, $i.e.$, La Niña consistently precedes the occurrence of a corresponding power outage (Figs. \ref{fig3}c and g). La Niña influences PONs by modulating the frequency of heatwave in the subsequent summer (Fig. \ref{fig3}d), and the frequency of extreme precipitations in the simultaneous winter (Fig. \ref{fig3}h). Specifically, in the summer, La Niña instigates a surge in heatwaves across the southern and mid-eastern U.S., while in the winter, La Niña amplifies the frequencies of extreme precipitations in the central and eastern U.S., consequently giving rise to the PONs.

\section*{Increasing heatwave and power outages in the future}

\begin{figure}[htb]
	\centering
	\includegraphics[width=0.9\textwidth]{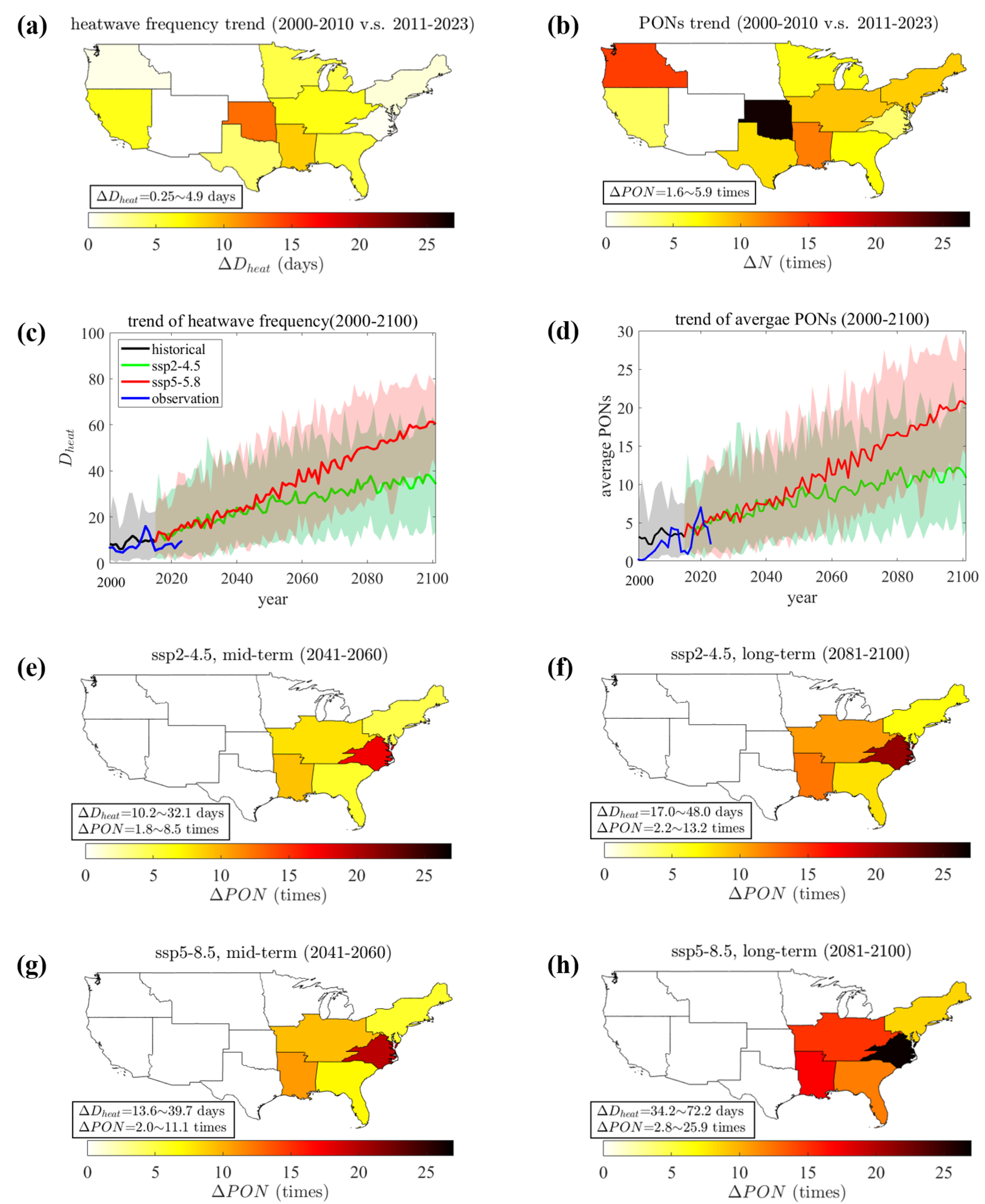}
	\caption{\textbf{Historical and future projected springtime PONs based on the changes in heatwave frequencies.} The amplified ratio of (a) heatwave frequencies and (b) PONs during the period 2000-2010 to the period 2011-2023. In (b), no weather-related power outage occurs in the climate region in black (S1) during 2000-2010. (c) Time series of observed and projected heatwave frequencies and (d) observed and estimated PONs in the spring from 2000 to 2100. SSP 2-4.5 and SSP 5-8.5 denote future moderate and high emission scenarios, respectively. The gray areas denote the upper and lower boundaries of historical heatwave frequencies and corresponding PONs during 2000-2014. From 2015 to 2100, the areas in green and red stand for the upper and lower boundaries of heatwave frequencies and corresponding PONs among the 18 CMIP6 GCMs under the SSP 2-4.5 and SSP 5-8.5 scenario, respectively. The lines in black, green, and red indicate the historical, SSP 2-4.5 scenario average, and SSP 5-8.5 scenario average heatwave frequencies and PONs. The blue lines represent the observed average heatwave frequencies and PONs. (e-f) Estimated amplified ratio of PONs under SSP 2-4.5 scenario. (g-h) Estimated amplified ratio of PONs under SSP 5-8.5 scenario. Color bars and text boxes in (e-h) indicate the amplified ratio in PONs and heatwave frequencies. Compared with historical (2000-2023) average PONs, (e) and (g) show the average amplified ratio in PONs for the mid-term future  (2041-2060), (f) and (h) for the long-term future (2081-2100).}\label{fig4}
\end{figure}

Heatwave is one of the leading factors for historical weather-related power outages (Figs. \ref{fig2}e and \ref{fig3}g). Over the past two decades, heatwaves prolonged across all climate regions, $i.e.$, the average heatwave frequencies during spring from 2011 to 2023 surpass those from 2000 to 2010 by $\Delta D_{heat}$ = 0.25 $\sim$ 4.9 days (Fig. \ref{fig4}a). Meanwhile, the average PONs also increased by $\Delta PON$ = 1.6 $\sim$ 5.9 times (Fig. \ref{fig4}b). In future, heatwaves are anticipated to be more frequent, persistent, and intense in almost all populated regions \cite{bib71}. 
We estimate a remarkable amplified trend of springtime PONs over the U.S. in the future, together with the projection of continuously growing heatwave frequencies projected by the CMIP6 (see Methods for details).

Surface temperature data from 18 CMIP6 General Circulation Models (GCMs, Supplementary Tab. 1) unveil the anticipated changes in heatwave frequencies (Fig. \ref{fig4}c). We identify that for spring from 2000 to 2100, there is a consistently increasing trend of heatwave frequencies on average among five climate regions in the eastern U.S. (OV, S2, SE1, SE2, and NE), where PONs are significantly correlated with heatwave frequencies (p$\le$0.05) during 2000 to 2023. By 2100 spring, the average heatwave frequencies will reach approximately 35 days under a moderate emission (SSP 2-4.5, see Methods for details) scenario, and nearly 60 days under a high emission (SSP 5-8.5, see Methods for details) scenario. For each climate region, a linear regression model is fitted by the historical (2000 to 2023) heatwave frequencies with corresponding PONs, on which the expected PONs in future are evaluated by projecting the heatwave frequencies from 2000 to 2100. The results in Fig. \ref{fig4}d show that if the power system maintains its current development status without upgrades tailored to heatwaves, the average PONs will increase by about 10 times (under SSP 2-4.5 scenario) to 20 times (under SSP 5-8.5 scenario) at the end of this century. 

Average PONs rise with the growing heatwave frequencies, yet the increased proportion of PONs varies across emission scenarios, periods, and climate regions (Figs. \ref{fig4}e-h). Under a moderate emission (SSP 2-4.5) scenario, compared to the historically average PONs from 2000 to 2023, PONs will increase about $\Delta PON$ = 1.8 $\sim$ 8.5 times by the mid-term (2041-2060) of the 21st century with increasing frequencies of heatwaves $\Delta D_{heat}$ = 10.2 $\sim$ 32.1 days (Fig. \ref{fig4}e), which is further anticipated to surge to about $\Delta PON$ = 2.2 $\sim$ 13.2 times by the long-term (2081-2100) of the 21st century with $\Delta D_{heat}$ = 13.6 $\sim$ 39.7 days (Fig. \ref{fig4}f). For the high emission (SSP 5-8.5) scenario, a more serious increase of PONs is observed, $i.e.$, $\Delta PON$ are projected to increase approximately $\Delta PON$ = 2 $\sim$ 11 times in the mid-term with $\Delta D_{heat}$ = 17 $\sim$ 48 days (Fig. \ref{fig4}g), with a substantial uptick of $\Delta D_{heat}$ = 2.8 $\sim$ 25.9 times in the long-term with $\Delta D_{heat}$ = 34.2 $\sim$ 72.2 days (Fig. \ref{fig4}h). From the climate region perspective, the PON in the SE1 region escalates most under given emission scenarios and phases, followed by the PONs in OV and S2 regions, while the PONs in SE2 and NE regions demonstrate a relatively lower growth.

\section*{Conclusion and discussion} 
Through unraveling the spatio-temporal correlation patterns between ENSO and weather-related PONs, we establish a nexus among the global ENSO, local extreme weather, and power outages. This study highlights a significant linkage between weather-related PONs in the U.S. and La Niña (the negative phase of ENSO) intensity. Specifically, power outages are prone to occur during the La Niña phase, and the stronger the La Niña, the more power outages happen. The underlying mechanism is that La Niña modulates the occurrences of extreme weather events, including heatwaves, cold snaps, and extreme precipitations, consequently triggering weather-related power outages across seasons and climate regions. We demonstrate that the power outages closely correlate with La Niña occurring primarily in the subsequent spring, summer, and the simultaneous winter for the western, southern, and mid-eastern U.S. This pattern aligns well with historical observations, $i.e.$, according to the 2022 U.S. Department of Energy, seven of the top 10 states with the most power outages over the past 20 years are located in the western, southern, or mid-eastern regions (California, Texas, Pennsylvania, Florida, Ohio, North Carolina, and Illinois), with PONs ranked by season as winter, spring, summer, and autumn \cite{bib70}. Furthermore, heatwave frequencies are projected to markedly increase during the 21st century. Compared with 2000 to 2023, we find the sustained rise in heatwave frequencies will amplify the average PONs by up to 11 times more by the mid-term (2041-2060), and even 25.9 times more by the long-term (2081-2100) of the 21st century.

Despite the intrinsic correlation between La Niña and weather-related power outages in the U.S. is significantly uncovered, our study has some limitations. First, we focus the scope of climate phenomena on ENSO. It is crucial to recognize that numerous other internal climate phenomena, such as PDO (Pacific Decadal Oscillation), NAO (North Atlantic Oscillation), IOD (Indian Ocean Dipole) $etc.$, also play critical roles in modulating local weather systems, which potentially affects the operation of power systems. For example, it was found that wind speeds of wind farms vary under different NAO states, yielding a difference in the average wind power output of up to 10\% \cite{bib23}. Second, only the PONs within the continental U.S. are studied. Research efforts are worth to be extended to other countries and regions as well. Further investigation of various climate phenomena and their impacts on power outages worldwide is necessary but complex, since different climate phenomena share intertwined effects on weather-related power outages across time and regions, the research of which is also hinged upon the data availability, volume, and quality from both earth science community and electrical industry \cite{bib24, bib31}.

Extreme and consecutive La Niña probably occur more frequently in this century \cite{bib68, bib69}, further exacerbating the risk of weather-related power outages. Moreover, as the penetration of renewable energies such as photovoltaics and wind power increases, the resilient operation of future power systems becomes more contingent upon the climate and weather \cite{bib25}. It is essential to develop climate-coupled system planning methods to enhance the power system resilience. Our findings of the spatio-temporal correlation pattern between La Nina and PONs provide complementary meteorological information for mid and long-term power system planning. Evidence shows that patterns of ENSO can be predicted some months in advance, which is often more successful than the common simulating short-term climate point data \cite{bib30}.

\section*{Methods} 

\textbf{Datasets}: In our study, four datasets are utilized, including power outage records, ENSO monitoring index, $in$ $situ$ observation-based weather data, and weather data generated by the CMIP6.

First, power outage records are collected from the Electric Emergency Incident and Disturbance Report (Form DOE-417) at the U.S. Department of Energy (DOE). We use the power outage records with power outage events spanning 24 years (from March 2000 to March 2023) in 48 states in the continental U.S. (excludes Hawaii and Alaska). Totally 1554 power outage events caused by severe weather are considered after the events with missing or mismatched information are filtered out. Note that the power outage records only involve major power outage events satisfying specific criteria, $e.g.$, loss of electric service to more than 50,000 customers for 1 h or more \cite{bib35}. The original power outage records have the attribute about the outage beginning date and time, the monthly PON is counted by the number of power outage events beginning in the given month. 

Second, to characterize the ENSO intensity, the monthly time series of five typical ENSO monitoring indices are used. The five indices are the Multivariate ENSO Index (MEI), Ni\~no 3 index, Ni\~no 4 index, Ni\~no 3.4 index, and Southern Oscillation Index (SOI) \cite{bib5, bib7, bib10}. 

Third, the $in$ $situ$ observation-based weather data refers to the daily 2m surface temperature and total precipitation from the Climate Prediction Center (CPC) global unified temperature and precipitation dataset, with $0.5^{\circ} \times 0.5^{\circ}$ horizontal resolution. The weather data ranges from March 2000 to March 2023, which is consistent with the studied period of power outage records.

Last, the future weather data is provided by the CMIP6. The CMIP6 is widely applied to compare different GCMs that project the climate change conditions with given Shared Socioeconomic Pathways (SSPs). The SSPs are climate change scenarios for characterizing future socioeconomic development up to the end of the 21st century, where the SSP 2-4.5 and SSP 5-8.5 scenarios represent the moderate and high greenhouse gas emission scenarios.

\textbf{Frequencies of extreme weathers}: Frequencies of extreme weathers refer to how many days extreme weathers occur. We consider three types of extreme weather: heatwaves, cold snaps, and extreme precipitation. The weather of a day is defined as the heatwave (cold snap) if its temperature is larger than (smaller than) a specified threshold, $i.e.$, a sliding window is set centered on that day with $\pm \Delta$ days sliding, the 95th percentile (5th percentile) of temperatures from the same period given by the sliding window among all studied years serves as the threshold. Similarly, the weather of a day is defined as extreme precipitation if its precipitation exceeds the corresponding threshold. The length of the sliding window is set as $\Delta = 15$ days.

\textbf{Data preprocessing}: Let the scalar time series $X = \left \{ x(t) \right \}$ be the monthly PON, ENSO monitoring index time series, or the $in$ $situ$ observation-based weather data at a given coordinate $(i,j)$, where $x(t)$ is the observation at month $t=1,2,...,T$ with $T$=288 (12 months$\times$24 years). An alternative representation of $X$ is $X = \left \{ x(m,n) \right \}$, where $m$ and $n$ denote the month and year ($m=1,2,..., 12$ and $n=1,2,...,24$). Without any specification in the main
text, the monthly PON series, the ENSO monitoring index series, and weather data are processed by the following three procedures.

Firstly, we remove the linear trend of $X$. The detrended result $\hat{X} = \left \{ \hat{x}(t) \right \}$ is calculated as:
\begin{subequations}
	\begin{align}
		x(t) = a + bt + \epsilon (t) \label{eq2a} \\
		\hat{x}(t) = x(t) - (a + bt) \label{eq2b}
	\end{align} \label{eq2}
\end{subequations} 
Eq. \ref{eq2a} is the linear regression model of $X$ with intercept $a$, slope $b$, and random error term $\epsilon (t)$, and $\hat{x}(t)$ in Eq. \ref{eq2b} denotes $x(t)$ after detrending. 

Secondly, the anomaly is calculated as $\grave{X} = \left \{ \grave{x}(m,n) \right \}$:
\begin{equation}
	\grave{x}(m,n)=\hat{x}(m,n) - \frac{1}{24} \sum_{n=1}^{24} \hat{x}(m,n)  \label{eq3}
\end{equation}
Eq. \ref{eq3} represents the anomaly at $n$th month among a given climatology (reference period). The climatology is set from 2000 to 2023, which is consistent with the period of power outage records in this study. The anomaly is used to indicate whether the observation for a given month is above or below the historical average for the same month.

Thirdly, the 3-month running mean is performed on $\grave{X}$, and we denote the result as $\tilde{X} = \left \{ \tilde{x}(t) \right \}$. For $t=2,3,...,287$, $\tilde{X}$ is calculated as:
\begin{equation}
	\tilde{x}(t) = \frac{1}{3}\left [ \grave{x}(t-1) + \grave{x}(t) + \grave{x}(t+1) \right ]    \label{eq4}
\end{equation}

In the main text, different subsets of $\tilde{X}$ under specific conditions could be used, $i.e.$, $\tilde{X}_{cond}=\left \{ \tilde{x}(m,n)\in X\mid C(\tilde{x}(m,n)) \right \}$, where $C(\cdot )$ is the condition function. For example, suppose $\tilde{X}$ is the time series of MEI, then the negative phase of MEI in spring is selected as $\tilde{X}_{cond}=\left \{ \tilde{x}(m,n)\in X\mid \tilde{x}(m,n)<0 \wedge m\in \left \{ 3,4,5 \right \}  \right \}$.

\textbf{Correlation Assessment}: The time delayed, cross-correlation (CC) function \cite{bib17} is applied to assess correlations between any two time series among the PON series, ENSO monitoring index, and weather data. Let $\tilde{X} =\left \{ \tilde{x}(t) \right \}$ and $\tilde{Y} =\left \{ \tilde{y}(t) \right \}$ be two time series among PON series, ENSO monitoring index, and weather data at a given coordinate $(i,j)$ after data processed by Eq. \ref{eq2} to Eq. \ref{eq4}, the time delayed CC function between $\tilde{X}$ and $\tilde{Y}$ is:
\begin{equation}
	R = \frac{ {\textstyle \sum_{t=1}^{T-k}}\left [ \tilde{x}(t+k) - \bar{x} \right ] \left [ \tilde{x}(t)-\bar{y}  \right ]  }{\sqrt{ {\textstyle \sum_{t=1}^{T}}\left [ \tilde{x}(t)-\bar{x}  \right ]^{2} {\textstyle \sum_{t=1}^{T}}\left [ \tilde{y}(t)-\bar{y}  \right ]^{2}  } } \label{eq5}
\end{equation}
where $\bar{x}$ and $\bar{y}$ represent the time average of $\tilde{X}$ and $\tilde{Y}$, and $k$ is the time lag. The maximum time leg in the main text is $\pm12$ months.

\textbf{ANOVA and Tukey's HSD test}: To test whether the PON during El Ni\~{n}o (EI), La Ni\~{n}a (LA), and neutral (NE) phases are significant different from each other, $e.g.$, PON during LA is significant more than that during EI, we adopt the statistical methods called Analysis of Variance (ANOVA) and Tukey's Honestly Significant Difference (HSD) test \cite{bib18}. According to phases of the EI, LA, and NE, we divide the PON series $\tilde{X}$ into three subsets, $i.e.$, $\tilde{X}^{EI}$, $\tilde{X}^{LA}$, and $\tilde{X}^{NE}$. The ANOVA is to verify whether means of all subsets are equal:
\begin{equation}
	H_{0}: \tau_{EI}= \tau_{LA}= \tau_{NE}= 0 \label{eq7}
\end{equation}
against the alternative
\begin{equation}
	H_{a}: \exists \tau \in \left \{ \tau_{EI}, \tau_{LA}, \tau_{NE}\right \} \text{is significantly different from 0} \label{eq8}
\end{equation}
where $\tau_{EI}$, $\tau_{LA}$, and $\tau_{NE}$ are the group effects, representing the differences in average PON during EI, LA, and NE phase, respectively, compared to the overall average PON. 

If the Null Hypothesis in Eq. \ref{eq7} is rejected, that is, the mean of at least one subset is significantly different from others, we will perform the HSD test to determine which pair of subsets share significant differences. A Tukey's HSD value can be calculated with mean square error (MSE) within each subset, MSE between pairs of subsets, and the given significance level  \cite{bib18}. Then we compare the difference in means between each pair of subsets to Tukey's HSD value. If the difference in means between two subsets is larger than Tukey's HSD value, then they are considered to have a significant difference.

\section*{Data availability}
The power outage data is available at the U.S. Department of Energy (DOE) website \url{https://www.oe.netl.doe.gov/oe417.aspx}. The ENSO monitoring index is provided by the National Centers for Environmental Information and National Weather Service (NOAA) at \url{https://psl.noaa.gov/data/climateindices/list}. The $in$ $situ$ observation-based weather data can be obtained from the Climate Prediction Center (CPC) global unified temperature and precipitation dataset at \url{https://psl.noaa.gov/data/gridded/data.cpc.globaltemp.html}. The future weather data is generated from the Coupled Model Intercomparison Project Phase 6 (CMIP6) simulations at \url{https://aims2.llnl.gov/search/cmip6/}.

\section*{Code availability}
The codes used to generate the results of this study are available on request from the authors.

\section*{Acknowledgements}
Long Huo would like to thank Jianxin Zhang, Zhen Su, Prof. Dieter Gerten, Mengke Wei, and Zhigang Wang for the helpful discussions of this study during Long Huo's visit at the Potsdam Institute for Climate Impact Research (PIK). Long Huo, Fenying Cai, and Kaiwen Li are supported by the China Scholarship Council (CSC) scholarship. This work is partially supported by the Youth Innovation Team of China Meteorological Administration“Climate change and its impacts in the Tibetan Plateau”(CMA2023QN16).

\newpage

\bibliography{sn-bibliography.bib}
\end{document}